\documentclass[a4paper,11pt]{article}
\usepackage{pos}
\usepackage{graphicx}

\newcommand{\manuscriptfigure}[2]{%
	\includegraphics[width=#2]{#1}%
}
\usepackage{amsmath}

\title{Reformulating Pfaffian Quantum Monte Carlo with the Hybrid Monte Carlo formalism}

\author*[a]{Thomas Hauschild}
\author[a]{Lin Wang}
\author[a]{Finn L. Temmen}
\author[a,b]{Thomas Luu}

\affiliation[a]{Institute for Advanced Simulation (IAS-4),\\
	Forschungszentrum Jülich, Germany}

\affiliation[b]{Helmholtz-Institut für Strahlen- und Kernphysik and Bethe Center for Theoretical Physics,\\
	Rheinische Friedrich-Wilhelms-Universität Bonn, Germany}

\emailAdd{t.hauschild@fz-juelich.de}
\emailAdd{l.wang@fz-juelich.de}
\emailAdd{f.temmen@fz-juelich.de}
\emailAdd{t.luu@fz-juelich.de}

\abstract{Pfaffian quantum Monte Carlo extends auxiliary field methods to anomalous fermionic Hamiltonians, opening access to pairing systems and Hubbard--Stratonovich decompositions that are unavailable in determinant formulations. We show that this flexibility can be combined with the global updates of Hybrid Monte Carlo (HMC), leading to an algorithm scaling as $\mathcal{O}(N_\tau L^3)$, similar to its determinant quantum Monte Carlo (DQMC) counterpart without pseudofermions. The central observation is that the molecular dynamics force depends only on the magnitude of the Pfaffian weight, which admits a determinant representation and leads to a simplified force expression. This construction also yields unequal-time Majorana correlators at essentially no additional cost. We demonstrate the method for the interacting Kitaev chain using the imaginary and symmetric Majorana hopping channel. The resulting noncommuting auxiliary-field bond operators are treated through an even–odd decomposition. Their block structure permits an analytic reduction to a sparse effective interaction matrix of bandwidth two and allows explicit force expressions. Results agree with exact diagonalization and resolve low-energy modes localized at opposite boundaries, together with their increasing overlap near the topological transition of the non-interacting system. These findings establish HMC as an efficient sampling framework for Pfaffian auxiliary-field simulations.}

\FullConference{The 43rd International Symposium on Lattice Field Theory (Lattice 2026)\\
	July 26 to August 1, 2026\\
	University of Maryland, College Park, USA\\}

\begin{document}
	\maketitle
	
	\section{Introduction}
	\label{sec:intro}
	
	Auxiliary field quantum Monte Carlo methods are widely used to study interacting lattice fermions. Among them, determinant quantum Monte Carlo (DQMC) is a particularly powerful and well-established approach. In its standard formulation, however, DQMC requires the particle number to be conserved for every auxiliary-field configuration. Consequently, even when the underlying Hamiltonian conserves particle number, Hubbard--Stratonovich (HS) transformations that violate this symmetry at the level of individual field configurations cannot be used. Pfaffian quantum Monte Carlo (PfQMC) removes this restriction \cite{Han2024PfQMC}. Existing PfQMC implementations have exclusively employed Blankenbecler--Scalapino--Sugar (BSS) sampling \cite{Blankenbecler1981}. Hybrid Monte Carlo (HMC) provides complementary advantages, including global updates, tunable acceptance rates, and favorable scaling when combined with pseudofermions \cite{HMC}.
	
	We formulate PfQMC within HMC while retaining the computational scaling of the corresponding DQMC algorithm. We achieve this by deriving an efficient expression for the molecular dynamics (MD) force that depends only on the magnitude of the Pfaffian weight. Majorana correlation functions arise as intermediate quantities in the force calculation and can therefore be accumulated at little additional cost. We demonstrate the method for the interacting Kitaev chain using the symmetric imaginary Majorana hopping channel. In this formulation, neighboring auxiliary-field bond operators do not commute, which we handle through an even--odd bond decomposition. We exploit the block-diagonal structure of the resulting interaction kernels to combine their exponentials into a sparse effective interaction matrix and simplify the evaluation of both the weight and the MD force. We validate the algorithm against exact diagonalization, confirm its predicted $\mathcal{O}(N_\tau L^3)$ scaling through runtime measurements, and use unequal-time correlators and localization measures to characterize Majorana edge modes.
	
	\section{Pfaffian formulation of auxiliary-field QMC}
	\label{sec:background}
	To represent pairing bilinears excluded from standard DQMC, PfQMC uses a $2L$-dimensional Majorana basis for $L$ fermionic sites \cite{Han2024PfQMC},
	\begin{equation}
		c_i=\tfrac{1}{2}\bigl(\gamma_i^{(1)}+i\gamma_i^{(2)}\bigr),\qquad
		c_i^{\dagger}=\tfrac{1}{2}\bigl(\gamma_i^{(1)}-i\gamma_i^{(2)}\bigr),\qquad
		\{\gamma_i^{(a)},\gamma_j^{(b)}\}=2\delta_{ij}\delta_{ab}.
		\label{eq:majorana_definition}
	\end{equation}
	After a Suzuki--Trotter decomposition and Hubbard--Stratonovich transformation, we write the partition function as
	\begin{align}
		Z
		&=\int \mathcal{D}\phi\,e^{-S_B[\phi]}
		\operatorname{Tr}\!\left\{
		\prod_{l=1}^{N_\tau}
		\exp\!\left(-\tfrac{1}{4}\vec{\gamma}^{\,T}{A}\vec{\gamma}\right)
		\exp\!\left(-\tfrac{1}{4}\vec{\gamma}^{\,T}{B}[\phi^l]\vec{\gamma}\right)
		\right\}
		= \int \mathcal{D}\phi\,e^{-S_B[\phi]}W_{\mathrm{Pf}}[\phi].
		\label{eq:pf_partition}
	\end{align}
	Here $N_\tau$ is the number of time slices, $\phi$ is the auxiliary field, $S_B$ is the bosonic part of the action, and the $2L\times2L$ kernels ${A}$ and ${B}[\phi^l]$ contain the field-independent and field-dependent one-body terms, respectively. With the inverse temperature $\beta = \frac{1}{T}$, where $T$ is the temperature, the width of a time slice is given by $\Delta_{\tau} = \frac{\beta}{N_\tau}$. 
	
	$W_{\mathrm{Pf}}[\phi]$ is the fermionic trace. Following Ref.~\cite{Han2024PfQMC}, we represent each imaginary-time factor as a pair of a $2L\times2L$ matrix and a scalar and compose these pairs. With $I$ the identity matrix and
	\begin{align}
		G(X) = \left(I - X\right)\left(I + X\right)^{-1}, \qquad
		\eta(X)=\operatorname{Pf}\!\left(\begin{smallmatrix}
			\sqrt{2}\sinh(X/4) & -I\\
			I & \sqrt{2}\sinh(X/4)
		\end{smallmatrix}\right),
		\label{eq:eta_definition}
	\end{align}
	two pairs compose as
	\begin{align}
		\left(e^{-X}, \eta\left(X\right) \right) \circ \left(e^{-Y}, \eta\left(Y\right) \right) = \left(e^{-X}e^{-Y},\, \eta\left(X\right)\eta\left(Y\right)\operatorname{Pf}\left[\begin{smallmatrix}
			G\left(e^{-X}\right) & -I \\
			I & G\left(e^{-Y}\right)
		\end{smallmatrix}\right] \right) ,
		\label{eq:composition_rule}
	\end{align}
	an associative operation. With the shorthand
	\begin{equation}
		{R}_{a\ldots b}=e^{-{A}}e^{-{B}[\phi^a]}\cdots
		e^{-{A}}e^{-{B}[\phi^b]},\qquad
		{M}\equiv{M}[\phi]=I+{R}_{1\ldots N_\tau},
		\label{eq:pf_transfer_definitions}
	\end{equation}
	the $2N_{\tau}$ factors of a configuration compose to
	\begin{align}
		\left(R_{1\ldots N_{\tau}},\, W_{\mathrm{Pf}}[\phi] \right)
		={}& \left(e^{-A}, \eta\left(A\right) \right)\circ \left(e^{-B[\phi^{1}]}, \eta\left(B[\phi^{1}]\right) \right)\circ \dots \nonumber \\
		&\circ \left(e^{-A}, \eta\left(A\right) \right) \circ \left(e^{-B[\phi^{N_{\tau}}]}, \eta\left(B[\phi^{N_{\tau}}]\right) \right),
		\label{eq:weight}
	\end{align}
	leaving the weight as the second component.
	
	A sign problem may appear, as the weight is generally neither real nor positive. The usual DQMC reweighting procedures apply. The computational cost is $\mathcal{O}\left(N_{\tau} L^{3} \right)$. The unequal-time Majorana correlator takes the form
	\begin{equation}
		\left\langle\gamma_i^{(a)}(l)\gamma_j^{(b)}(0)\right\rangle
		=Z^{-1}\int\mathcal{D}\phi\,e^{-S_B[\phi]}W_{\mathrm{Pf}}[\phi]\,
		\left[2{M}^{-1}{R}_{1\ldots l}\right]_{(a,i)(b,j)} .
	\end{equation}
	
	\section{Hybrid Monte Carlo for PfQMC}
	\label{sec:hmc}
	The computational cost of HMC sampling is dominated by two operations: evaluating the weight itself and repeatedly computing the derivative of its logarithm along the MD trajectory. For a detailed explanation of HMC, the reader is directed to Ref.~\cite{HMC}. In the following section, stabilization will not be addressed. For later simulations, stabilization methods presented in \cite{LuuEtAl2026, Bauer2020} are reformulated and employed for PfQMC. We also use radial updates to ensure ergodicity \cite{Temmen2025}.
	\subsection{Efficient force evaluation}
	\label{sec:force}
	Direct differentiation of the Pfaffian weight is computationally expensive. HMC sampling, however, requires only the real part of the logarithmic derivative, which can be obtained from the magnitude of the weight. The magnitude of the Majorana trace satisfies $\left|W_{\mathrm{Pf}}[\phi]\right|=\left|\sqrt{\det{M}[\phi]}\right|$. The gradient of the real part of the action $S_{R}$ is therefore
	\begin{align}
		\frac{\partial S_R}{\partial\phi_x^l}
		&=\frac{\partial S_B}{\partial\phi_x^l}
		-\frac{1}{2}\operatorname{Re}\!\left[
		\operatorname{Tr}\!\left\{
		\frac{\partial}{\partial\phi_x^l}
		\left(e^{-{A}}e^{-{B}[\phi^l]}\right)
		{R}_{l+1\ldots N_\tau}{M}^{-1}{R}_{1\ldots l-1}
		\right\}\right].
	\end{align}
	For an interaction kernel of the form
	\begin{equation}
		{B}[\phi^l]_{(f,i)(f',j)}
		=r\,\delta_{ij}\phi_j^l
		\left(\delta_{f1}\delta_{f'2}-\delta_{f2}\delta_{f'1}\right),
		\label{eq:pf_diagonal_interaction}
	\end{equation}
	where $f,f'\in\{1,2\}$ label the two Majorana components and $r$ is a complex number, the determinant contribution to the force becomes
	\begin{align}
		\frac{\partial \left(S_R-S_B\right)}{\partial\phi_x^l} = \frac{1}{2}\operatorname{Re}\!\left\{r\left(\left[{R}_{l+1\ldots N_\tau}{M}^{-1}{R}_{1\ldots l}\right]_{(1,x)(2,x)}-\left[{R}_{l+1\ldots N_\tau}{M}^{-1}{R}_{1\ldots l}\right]_{(2,x)(1,x)}\right)\right\}. \label{eq:pf_force}
	\end{align}
	This expression is the Majorana analog of the diagonal-interaction result in DQMC. Other choices for common decoupling channels will yield similar expressions. Unequal-time correlators appear as intermediate quantities in the force calculation and can be accumulated with negligible additional cost, similar to the DQMC formulation in Ref.~\cite{LuuEtAl2026}.
	\subsection{Evaluation of the weight}
	Because we require the phase as well as the magnitude of the weight, we cannot reuse the determinant representation of its magnitude. Instead we need to find an efficient procedure to evaluate Eq.~(\ref{eq:weight}). Since the composition rule is associative, the evaluation can be accelerated substantially by parallelizing it. Further, while it is usually not possible to find closed form expressions of $G(e^{-A}e^{-B[\phi^1]}...e^{-A}e^{-B[\phi^t]})$, in many cases it is possible to find them for $G(e^{-B[\phi^t]})$, which further accelerates the evaluation of the weight. Often it is also possible to find simple closed form expressions for $\eta \left(B[\phi^t]\right)$. The main computational cost comes from evaluating the matrix products required for the Pfaffians. To increase stability of the algorithm, it is also advisable to evaluate the weight as a sum of logarithmic expressions. 
	
	\section{Application to the interacting Kitaev chain}
	\label{sec:kitaev}
	We demonstrate the method for the interacting Kitaev chain, a one-dimensional model of spinless fermions with nearest-neighbor hopping, $p$-wave pairing, and nearest-neighbor density interactions \cite{Kitaev2001}. For open boundary conditions, the Hamiltonian is
	\begin{align}
		H&=\sum_{j=1}^{L-1}\Bigl[
		-t\,c_j^{\dagger}c_{j+1}
		+\Delta\,c_{j+1}^{\dagger}c_j^{\dagger}
		+\mathrm{H.c.}
		+V\bigl(n_j-\tfrac{1}{2}\bigr)\bigl(n_{j+1}-\tfrac{1}{2}\bigr)
		\Bigr]
		-\mu\sum_{j=1}^{L}n_j,
		\label{eq:kitaev_hamiltonian}
	\end{align}
	where $n_j=c_j^{\dagger}c_j$, $V$ is the density--density coupling, $t$ the hopping amplitude, $\Delta$ the pairing amplitude, and $\mu$ the chemical potential. The coexistence of explicit pairing and interactions makes this model a natural test of the PfQMC formulation.

	With an appropriate gauge convention, the Majorana representation of the Hamiltonian is
	\begin{align}
		H={}&\sum_{j=1}^{L-1}\Bigl\{
		\tfrac{i}{2}\left[(-1)^j t-\Delta\right]
		\gamma^{(1)}_j\gamma^{(1)}_{j+1}
		+\tfrac{i}{2}\left[(-1)^j t+\Delta\right]
		\gamma^{(2)}_j\gamma^{(2)}_{j+1}
		\notag\\
		&\qquad
		+\tfrac{V}{4}
		\gamma^{(1)}_j\gamma^{(1)}_{j+1}
		\gamma^{(2)}_j\gamma^{(2)}_{j+1}
		\Bigr\}
		-\tfrac{i\mu}{2}\sum_{j=1}^{L}
		\gamma^{(1)}_j\gamma^{(2)}_j .
		\label{eq:kitaev_majorana}
	\end{align}
	For the simulations reported here, we use the symmetric and imaginary Majorana hopping channel, which follows from the identity $\gamma_j^{(1)}\gamma_{j+1}^{(1)}\gamma_j^{(2)}\gamma_{j+1}^{(2)}
	=\frac{1}{2}\Lambda_j^{2}+\text{const.}$, with $\Lambda_j:=\gamma_j^{(1)}\gamma_{j+1}^{(1)}+\gamma_j^{(2)}\gamma_{j+1}^{(2)}$. This choice alleviates the sign problem in the vicinity of $\mu=\Delta=0$, but it also complicates the structure of the interaction matrices. While $\left[\Lambda_j^{2},\Lambda_k^{2}\right]=0$, the same does not hold configuration-wise after performing the HS transformation, where $\left[\Lambda_j,\Lambda_k\right]\neq0$. The interactions can therefore not be rewritten on a single exponential. All even (odd) bond interactions do commute among themselves however, so grouping together all even bonds and all odd bonds allows us to write the interactions onto two separate exponentials. A single time slice is then handled by the kernels
	\begin{align}
		A_{(f,k)(f',j)}&=i\Delta_{\tau}\left[\Delta\left(\sigma_{z}\right)_{ff'}\delta_{\langle kj\rangle}(k-j)-t\,\delta_{ff'}\delta_{\langle kj\rangle}(-1)^{k}-i\mu\left(\sigma_{y}\right)_{ff'}\delta_{kj}\right],\\
		B^{\mathrm{odd}}[\phi^{l}]_{(f,k)(f',j)}&=-2i\,\delta_{ff'}\Bigl[\,\textstyle\bigoplus_{i\ \text{odd}}\left(\begin{smallmatrix}
			0 & \phi_{i}^{l}\\ -\phi_{i}^{l} & 0
		\end{smallmatrix}\right)\,\left(\oplus\,0\right)_{\text{if }L\text{ odd}}\Bigr]_{kj},\\
		B^{\mathrm{even}}[\phi^{l}]_{(f,k)(f',j)}&=-2i\,\delta_{ff'}\Bigl[\,0\oplus\textstyle\bigoplus_{i\ \text{even}}\left(\begin{smallmatrix}
			0 & \phi_{i}^{l}\\ -\phi_{i}^{l} & 0
		\end{smallmatrix}\right)\,\left(\oplus\,0\right)_{\text{if }L\text{ even}}\Bigr]_{kj}.
	\end{align}
	Due to the block-diagonal structure of the interaction matrices, many objects required for the algorithm can be calculated analytically. This lets us effectively reduce the number of kernels per time slice from three back to two. For the weight we find
	\begin{align}
		W_{\mathrm{Pf}}[\phi]&=\left[\prod_{t=1}^{N_{\tau}}\prod_{i=1}^{L-1}\cosh^{2}\left(\phi^{t}_{i}\right)\right]\prod_{m=1}^{2N_{\tau}-1}\operatorname{Pf}\begin{bmatrix}
			G_{1\ldots m} & -I\\
			I & G_{m+1}
		\end{bmatrix},\label{eq:weight_xi}\\
		G_{1\ldots m}&:=\left(I-M_{1}\cdots M_{m}\right)\left(I+M_{1}\cdots M_{m}\right)^{-1},\label{eq:caterpillar_G}\\
		\left(M_{1},\ldots,M_{2N_{\tau}}\right)&:=\left(e^{-A},\xi^{1},\ldots,e^{-A},\xi^{N_{\tau}}\right),\label{eq:caterpillar_M}
	\end{align}
	where we have defined the new effective interaction matrix $\xi^{t}:=e^{-B^{\mathrm{odd}}[\phi^{t}]}e^{-B^{\mathrm{even}}[\phi^{t}]}$. We find $\xi^{t}=\operatorname{diag}\left(\alpha^{t},\alpha^{t}\right)$, where $\alpha^{t}$ is an $L\times L$ matrix whose only non-vanishing entries are
	\begin{equation}
		\begin{array}{@{}l@{\quad}r@{\,}c@{\,}l@{\qquad}r@{\,}c@{\,}l@{}}
			m\ \text{odd}:  & \alpha^{t}_{m,m-1}&=&-ic_{m}s_{m-1},  & \alpha^{t}_{m,m}  &=&c_{m}c_{m-1},\\[1pt]
			                & \alpha^{t}_{m,m+1}&=&is_{m}c_{m+1},   & \alpha^{t}_{m,m+2}&=&-s_{m}s_{m+1},\\[3pt]
			m\ \text{even}: & \alpha^{t}_{m,m-2}&=&-s_{m-1}s_{m-2}, & \alpha^{t}_{m,m-1}&=&-is_{m-1}c_{m-2},\\[1pt]
			                & \alpha^{t}_{m,m}  &=&c_{m-1}c_{m},    & \alpha^{t}_{m,m+1}&=&ic_{m-1}s_{m}.
		\end{array}
		\label{eq:alpha_entries}
	\end{equation}
	Here, we have defined $c_{m}=\cosh\left(2\phi^{t}_{m}\right)$ and $s_{m}=\sinh\left(2\phi^{t}_{m}\right)$. The fields are padded by $\phi^{t}_{0}=\phi^{t}_{L}:=0$, so that $c_{0}=c_{L}=1$ and $s_{0}=s_{L}=0$. With this convention every entry whose column index leaves the range $1,\dots,L$ carries a factor $s_{0}$ or $s_{L}$ and vanishes, so the site belonging to no bond of a given parity needs no separate treatment. This new effective interaction matrix is still sparse and has bandwidth two.
	
	This analytical reduction also lets us simplify the evaluation of the gradient,
	\begin{equation}
		\frac{\partial S_{R}}{\partial\phi^{t}_{m}}=\frac{\partial S_{B}}{\partial\phi^{t}_{m}}-\frac{1}{2}\operatorname{Re}\operatorname{Tr}\left\{\left(\xi^{t}\right)^{-1}\frac{\partial\xi^{t}}{\partial\phi^{t}_{m}}N^{t}\right\},\qquad N^{t}:=R_{t+1\ldots N_{\tau}}M^{-1}R_{1\ldots t},
		\label{eq:log_force}
	\end{equation}
	where $R_{a\ldots b}$ and $M$ are those of Eq.~(\ref{eq:pf_transfer_definitions}) with $e^{-B[\phi^{t}]}$ replaced by $\xi^{t}$. Carrying out the trace gives
	\begin{align}
		\frac{\partial S_{R}}{\partial\phi^{t}_{m}}&=\frac{4\phi^{t}_{m}}{\Delta_{\tau}V}+\operatorname{Im}\sum_{f}\left(N^{t}_{(f,m+1)(f,m)}-N^{t}_{(f,m)(f,m+1)}\right),&&m\ \text{even},\label{eq:force_even}\\
		\frac{\partial S_{R}}{\partial\phi^{t}_{m}}&=\frac{4\phi^{t}_{m}}{\Delta_{\tau}V}+\operatorname{Im}\sum_{f}\left(\tilde{N}^{t}_{(f,m+1)(f,m)}-\tilde{N}^{t}_{(f,m)(f,m+1)}\right),&&m\ \text{odd},\label{eq:force_odd}
	\end{align}
	where $\tilde{N}^{t}:=e^{-B^{\mathrm{even}}[\phi^{t}]}N^{t}e^{B^{\mathrm{even}}[\phi^{t}]}$. Since the unequal-time correlators appear as intermediate results in the evaluation of $N^{t}$, they are again supplied at no additional cost.
		
	\section{Results}
	\label{sec:results}
	We apply PfQMC--HMC to the interacting Kitaev chain using the non-diagonal effective interaction matrix introduced above. We first give results validating the algorithm and then study the low-energy physics of the system.
	\begin{figure}[htbp]
		\centering
		\manuscriptfigure{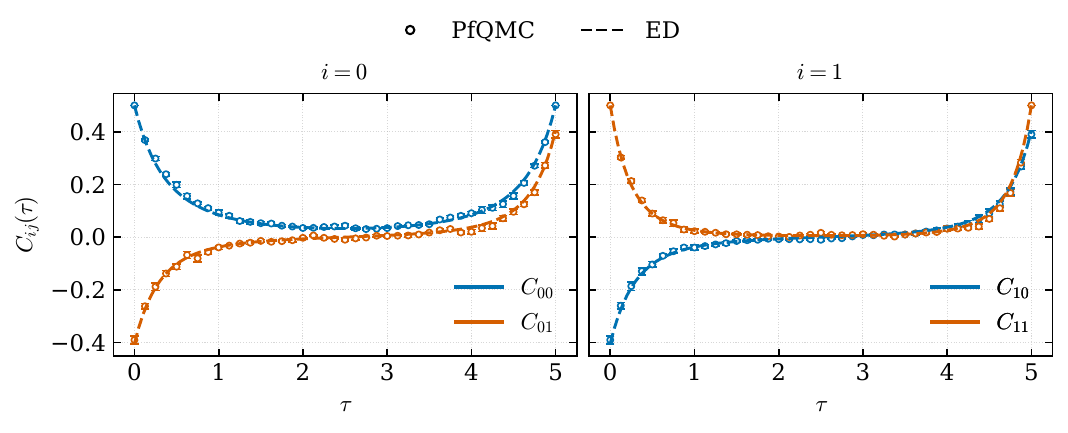}{0.95\textwidth}
		\caption{Comparison of PfQMC and ED results for the unequal-time Dirac correlators $C_{ij}(\tau)=\langle c_i(\tau)c_j^{\dagger}(0)\rangle$ in an interacting Kitaev chain with $L=6$ and inverse temperature $\beta=5$. The model parameters are $\mu=0$, $\Delta=1$, $V=4$, and $t=1$. The PfQMC parameters are $N_\tau=40$, $\Delta_\tau=0.125$, $N_{\mathrm{MD}}=7$ and $t_{\mathrm{MD}}=0.8$.}
		\label{fig:ed_comp}
	\end{figure}
	\subsection{Validation of the algorithm}
	Figure~\ref{fig:ed_comp} compares the PfQMC--HMC correlators with ED for a parameter set with a manageable average sign of $\langle \Sigma \rangle = 0.87 \pm 0.06$. The two methods agree within statistical uncertainties. We find comparable agreement across other parameter regimes we examined.
	
\begin{figure}[htbp]
		\centering
		\manuscriptfigure{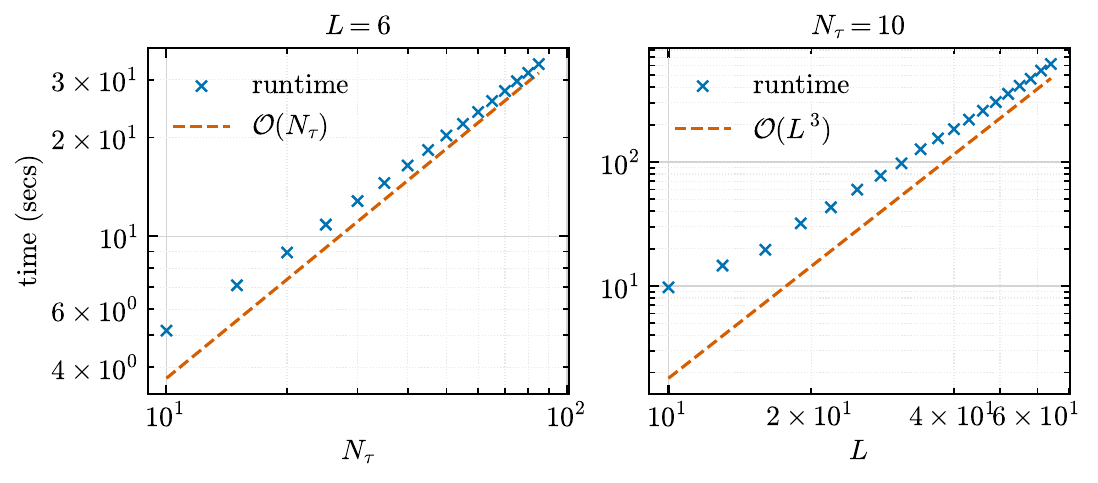}{0.95\textwidth}
		\caption{Runtime as a function of the number of time slices (left) and the system size (right). The parameters are $\mu=0.0$, $\Delta=1$, $t=1$, $V=1$, $t_{\mathrm{MD}}=0.8$, $N_{\mathrm{MD}}=7$, and 500 configurations. When varying the number of time slices, $\beta=5$, $\Delta_\tau=\beta/N_\tau$, and $L=6$. When varying the system size, $N_\tau=10$ and $\Delta_\tau=0.3$, so that $\beta=3$. The reference scalings have arbitrary coefficients.}
		\label{fig:scaling}
	\end{figure}

	Additionally, we determine the computational scaling of the algorithm by measuring the runtime for a fixed number of Monte Carlo updates while varying either the number of time slices or the system size. The results in Fig.~\ref{fig:scaling} reproduce the predicted asymptotic scaling: the runtime is linear in $N_\tau$ and cubic in $L$.

	\subsection{Majorana zero modes (MZMs)}
\begin{figure}[htbp]
		\centering
		\manuscriptfigure{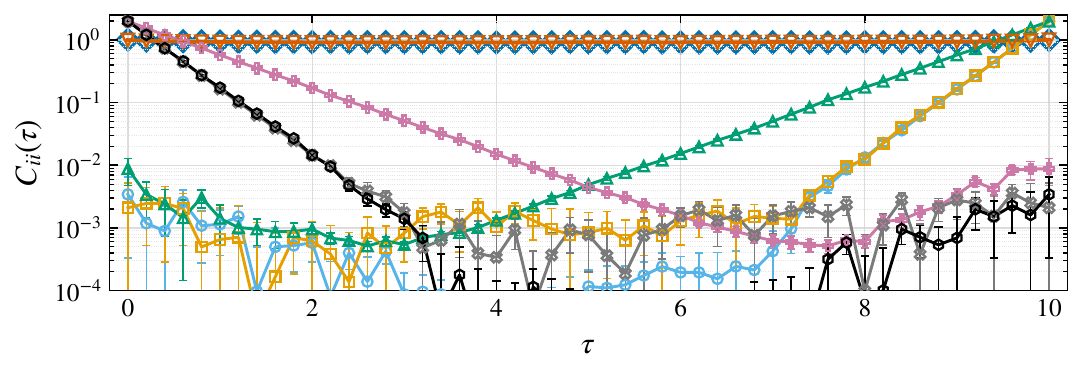}{0.95\textwidth}
		\caption{Unequal-time Majorana correlators $\langle\gamma_i(\tau)\gamma_i(0)\rangle$ in the eigenbasis of the non-interacting ${A}$ kernel for the eight lowest-energy modes. Excitation energies are encoded in the decay rates and can be extracted from the slopes on the logarithmic scale. The parameters are $L=14$, $t=1$, $V=1.0$, $\Delta=0.8$, $\Delta_\tau=0.125$, $N_\tau=60$, $N_{\mathrm{MD}}=7$ and $t_{\mathrm{MD}}=0.6$, with 3500 configurations.}
		\label{fig:mom_proj}
	\end{figure}

	We next consider an interacting chain with $L=14$. The low-energy spectrum is probed through unequal-imaginary-time correlators of operators transformed into the eigenbasis of the non-interacting ${A}$ kernel \cite{Luu2016}. Their decay rates encode the excitation energies. In Fig.~\ref{fig:mom_proj}, the rapidly decaying (or increasing) correlators correspond to excitations above the gap, whereas the nearly constant correlators identify a low-energy excitation. A zero eigenenergy does not yet prove the existence of an MZM.
	
	To characterize the low-energy excitation, we compute the single-site localization measure defined in Ref.~\cite{Samuelson2026}. MZMs are localized at opposite edges and have little spatial overlap. The measure can be inferred from the matrix elements $\langle e|\gamma_i^{(1)}|o\rangle$ and $\langle e|\gamma_i^{(2)}|o\rangle$, where \(|e\rangle\) and \(|o\rangle\) denote the two low-energy states below the excitation gap. These states belong to the even- and odd-parity sectors, respectively, as indicated by their labels \cite{Kitaev2001,Samuelson2026}. These matrix elements are not directly accessible in auxiliary-field QMC, but unequal-imaginary-time correlators provide proportional quantities:
	\begin{equation}
		\left\langle\gamma_i^{(a)}(\beta/2)\Gamma^0\right\rangle
		=2Z^{-1}\operatorname{Re}\!\left(
		\langle e|\gamma_i^{(a)}|o\rangle
		\langle o|\Gamma^0|e\rangle
		\right)
		e^{-\frac{\beta}{2}\left(\epsilon_o+\epsilon_e\right)}
		+\text{higher-energy terms},
	\end{equation}
	where $\epsilon_e$ and $\epsilon_o$ are the energies of the two low-energy states and $\Gamma^0$ is a non-interacting Majorana wave function. For a finite gap, the higher-energy terms are exponentially suppressed with increasing $\beta$. The second matrix element is independent of the site index $i$, so with an appropriate choice of the gauge, the correlator is proportional to the real part of the desired matrix element. The imaginary part can be calculated similarly from the counterpart of $\Gamma^0$. With this the localization measure can be calculated up to the normalization, which is not necessarily one, since many-particle contributions are not captured in the localization measure.
	\begin{figure}[htbp]
		\centering
		\manuscriptfigure{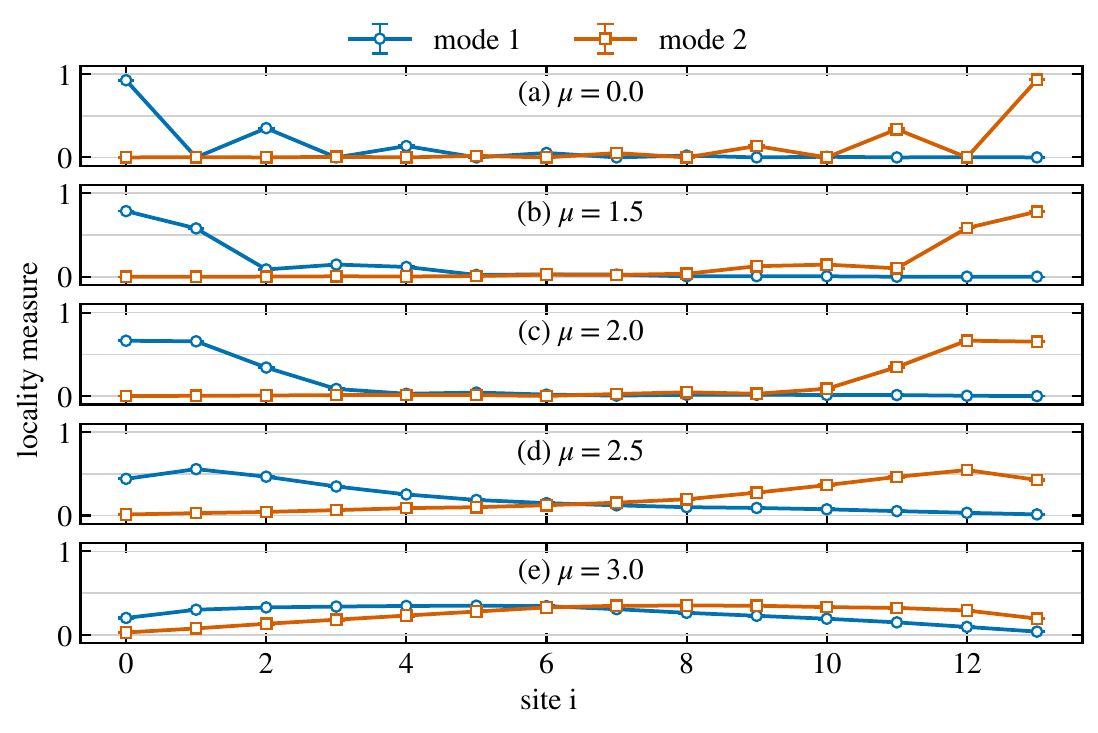}{0.80\textwidth}
		\caption{Unnormalized profiles of the single particle localization measure for several chemical potentials. The parameters are $L=14$, $t=1$, $V=1.0$, $\Delta=0.8$, $\Delta_\tau=0.2$, $N_{\tau}=50$, $N_{\mathrm{MD}} = 6$ and $t_{\mathrm{MD}}=0.6$. Error bars are mostly obscured by data markers.}
		\label{fig:loc_prof}
	\end{figure}
	In the non-interacting system, a phase transition occurs at $\mu=2t$, separating the topological phase for $\mu<2t$ with Majorana edge modes from the trivial phase for $\mu>2t$. We now investigate how this behavior is modified by interactions. Figure~\ref{fig:loc_prof} shows the localization measures for several chemical potentials in the interacting system. For $\mu<2$ (with $t=1$), the modes are clearly localized at opposite ends of the chain, indicating MZMs in this interacting parameter regime. Increasing $\mu$ shifts weight toward the center and increases the overlap. Near $\mu=2$, corresponding to the transition point of the non-interacting model, the overlap becomes large. Thus, the interacting results show qualitatively similar behavior to the non-interacting case.
	
	\section{Summary and outlook}
	\label{sec:conclusion}
	We formulated Pfaffian quantum Monte Carlo within HMC for systems with explicit or HS-induced particle-number violation. A determinant representation of the Pfaffian magnitude yields a DQMC-like force, unequal-time Majorana correlators at negligible additional cost, and asymptotic complexity $\mathcal{O}(N_\tau L^3)$.
	
	For the Kitaev chain with the imaginary symmetric Majorana hopping channel, the interaction operators cannot easily be rewritten on a single exponential, because configuration-wise the interaction operators no longer commute. Therefore we end up with three matrices per time slice as opposed to the usual two interaction matrices. By using analytical results for these matrices we could reduce the number of factors back to two by introducing a new effective interaction matrix of bandwidth two.
	The algorithm is consistent with other methods and can be used to identify MZMs \cite{Miao2017,Han2024PfQMC}.
	
	In future work we would like to incorporate pseudofermions. That would allow us to substantially increase performance, allowing the simulations of much larger system sizes \cite{Fucito1981}. It could also prove interesting to study additional decoupling channels and their effects on the sign problem. With the general method being established, the treatment of further, more realistic models is an exciting possibility.

	\acknowledgments
	This work was supported by the Deutsche Forschungsgemeinschaft (DFG, German Research Foundation) in part through the Cluster of Excellence `Color meets Flavor' (EXC 3107) and in part through the CRC 1639 NuMeriQS – Project number 511713970. We gratefully acknowledge the computing time granted by the JARA Vergabegremium and provided on the JARA Partition part of the supercomputer JURECA at Forschungszentrum Jülich \cite{JURECA2021}. Special thanks go to Johann Ostmeyer and Petar Sinilkov for numerous helpful discussions.

\end{document}